# Analysis of Network Coding in a Slotted ALOHA-based Two-Way Relay Network

A. Mahdavi Javid, M. Setayesh, F. Farhadi, and F. Ashtiani

Advanced Communications Research Institute (ACRI), Department of Electrical Engineering, Sharif University of Technology,

Emails: mjavid_a@ee.sharif.edu, setayesh_mehdi@ee.sharif.edu, farhadi@ee.sharif.edu, ashtianimt@sharif.edu

**Abstract**—This paper deals with a two-way relay network (TWRN) based on a slotted ALOHA protocol which utilizes network coding to exchange the packets. We proposed an analytical approach to study the behavior of such networks and the effects of network coding on the throughput, power, and queueing delay of the relay node. In addition, when end nodes are not saturated, our approach enables us to achieve the stability region of the network in different situations. Finally, we carry out some simulation to confirm the validity of the proposed analytical approach.

## I. INTRODUCTION

Network coding as an emerging technology has been turned out to be an efficient method to enhance the performance of wireless networks. Transmitting the combination of two packets instead of two individual packets not only increases the throughput of the network, but also decreases the transmission power at each node as well [1]–[4]. Meanwhile, the receiver node can decode the combined packet simply by using the packets it has stored through *opportunistic listening* [4].

A two-way relay network (TWRN), which consists of one relay node and two end nodes, is one of the most commonly used network structures in which the advantages of network coding have been extensively studied [5], [6]. TWRN structure can be considered as the basic structure in modern communication network scenarios, e.g., device-to-device (D2D) communication, WBAN, IoT, etc.

The main issue with which many of the current research make deal is the randomness of the arrival traffic into the buffers of each wireless node. In other words, the packets from two different sources may not be present at the same time in the buffer of the relay node. Therefore, in order to be able to combine two packets with different destinations, the relay node may have to wait a long time for the expected packet to arrive, due to this random behavior of the incoming traffic, which induces a could-be large delay in the network. In this situation, neither waiting a long time nor sending the uncoded (native) packet immediately is an appropriate solution for the relay node; hence, there seems to be a tradeoff between the end-to-end delay on the one hand and the power consumption as well as the throughput on the other hand. Such a transmission scheme for transmitting the packets is the one referred to as *opportunistic network coding* [7], [8].

There are some other works on the gain of network coding in TWRN in terms of the various network parameters such as throughput, delay, power consumption, etc, with a focus on Markov chains. Buffer-aware network coding was addressed in [9] which has employed a Markov model for representing the buffer states in order to analyze the power consumption and average delay in a network based on FDMA. Goseling *et al.* [10] used a two-dimensional continuous-time Markov chain to achieve analytical upper and lower bound on the energy consumption and the delay, while simply assumes separate arrival rates for each buffer in the relay node and therefore, there is no collision in the network. [11] proposed a coded packet priority access (CPPA) in which the coded packets have higher transmission probability than the native packets, and assumes the relay node transmits the coded packet successfully whenever there is a coding opportunity which simplifies the analysis of the Markov chains considerably. Umehara *et al.* [12] derived the achievable region of the throughput in a slotted ALOHA system, while [13] finds the maximum network throughput as a function of the number of the flows in the network by using a Markov model. Also, the authors in [14] derived the throughput in a slotted ALOHA system but under the assumption that the packet lengths are not necessarily the same. Umehara *et al.* utilized wireless network coding with unbalanced traffic, and proved that NC system has a better performance than non-NC system in slotted ALOHA system [15]. All these works [9]–[14] assumes that the end nodes of the TWRN are saturated—each of them has always some packets to transmit—which reduces the dimensionality of the Markov chain of the network. Nevertheless, this assumption loses its validity when the packet arrival fails due to lossy characteristic of the network, for example, noise.

Under the situation that the end nodes are unsaturated, end node buffers can be empty sometimes which, in turn, affects the departure rate of the other end node because the end nodes are interacting. One of the pioneer works regarding the interacting queues is Rao and Ephremides [16]. They obtained the stability region of a single-hop network in a slotted ALOHA system in which a collision can happen between two interacting queues. Krikidis [17] derived an inner bound for the stability region of a multiple-access broadcast decode-and-forward (MABC-DF) bidirectional relay channel, assuming that the two sources generate only bursty traffic. Recently, Amerimehr and Ashtiani [18] provided an analytical study of the average delay and throughput in a two-way relay network, which uses network coding to exchange the packets. They also derived the stability regions of three proposed opportunistic network coding schemes; but did not address the collisions that may occurred in the system between the relay node and the two end nodes which increases the complexity of the problem considerably.

In this paper, we provide an analytical approach to study the behavior of a slotted ALOHA-based TWRN, where the relay node uses three different transmission probabilities based on the status of its buffers. We have analyzed this network in two different situations: when the end nodes are saturated and when the end nodes are unsaturated. In each case, we have also addressed the effects of the collisions on the throughput, transmission power, and queueing delay of the relay node. In this way, our work can be partly considered as an extension to [15] which assumed a single transmission probability for the relay node to simplify the analysis. We apply the method introduced in [19] for analyzing a multi-buffered systems which paves our way to deal with more general situations of the network. The main contribution of this paper can be summarized as follows:

(1) We studied the effects of utilizing three different transmission probabilities for the relay node in the case of saturated end nodes, in terms of throughput, transmission power, and queueing delay.

(2) When the end nodes are unsaturated, in addition to transmission power and queueing delay, we have also presented the stability region of the network in different situations.

(3) We proposed an analytic approach to study interacting multi-buffer systems and it is shown to have the sufficient accuracy to assess the network performance.

The rest of the paper is arranged as follows; Section II illustrates the network scenarios in two situations: with saturated end nodes and with unsaturated end nodes. Section III describes the modeling of the system with saturated end nodes along with an analytical approach to find the throughput, the queueing delay, and the transmission power of the relay node. In Section IV, we present some numerical results to validate the proposed approach in the previous section. Sections V and VI include modeling of the system with unsaturated end nodes and the corresponding results, respectively. Finally, Section VII concludes the paper.

## II. NETWORK SCENARIO

Assume that two wireless nodes want to transmit their packets to each other, but because of a physical obstacle, they cannot transmit directly; therefore, they send their packets via a relay node R, as shown in Fig. 1. As indicated in Fig. 1, each end node can also represent the relaying traffic of several nodes under coverage of an access point (AP), sent to one of the nodes under coverage of another access point. Therefore, we can assume that each end node is a set of small nodes with different traffics. The relay node only receives packets from each end node and transmits them to the other end node and does not generate any traffic itself. Operating in one frequency channel and assuming that nodes use half-duplex (HD) transmission, each node cannot send and receive simultaneously. The protocol that is employed in this two-way relay network is Slotted Aloha, where collisions can occur for transmitted packets both in uplink and downlink. The end nodes may be saturated or unsaturated; in each case, we are interested in finding throughput, queueing delay, and transmission power of the relay node. We also assume interference and transmission range are such that the transmitted signal at each end node has negligible interference at the other one. For convenience, we assume constant packet length and that each node can send a packet and receive its ACK within the same time slot. In the case of sending a coded packet, we employ the simple ACK scheme proposed by Umehara [15] in which if a coded packet is successfully transmitted from the relay node, two ACK packets are transmitted from both destinations in a TDMA manner and within the same slot.

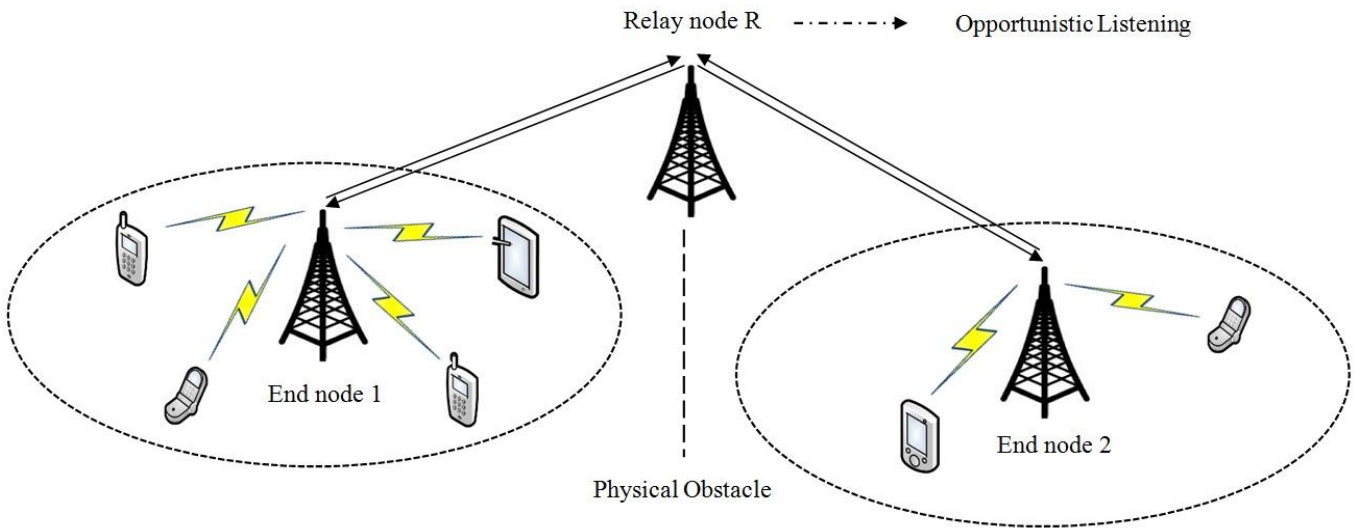

Fig. 1. Two-way wireless relay network with one relay node between the two end nodes

In the TWRN considered in this paper, the three nodes contend with each other in order to send their packets. Although the packets of each end node are eventually destined to the other end node, end node 1 or 2 sends their packets toward the relay node R. Then, the relay node sends the received packets toward their destinations. To this end, we assume that the relay node has two virtual buffers 1 and 2 corresponding to the sources of the received packets, as shown in Fig. 2. When the relay node receives a packet from end node $i$, stores it in the virtual buffer $i$ and tries to send it in a next slot. If both virtual buffers have some packets to send, node R combines the head-of-line packets of both virtual buffers by using bitwise XOR and transmits the coded packet with probability $q$. If only virtual buffer $i$ is nonempty and the other one is empty, node R transmits the native packet (i.e., uncoded packet) with probability $q_i$. We also assume that each end node $i$ has a transmission probability $g_i$. This scheme is similar to the one proposed in [15], except that we have different transmission probabilities, $q_i$, for sending the native packets of two virtual buffers. When node R combines the packets of both virtual buffers and sends the coded packet, the transmission power is obviously one-half that of sending the native packets separately. Moreover, throughput of the network equals to the arrival rate at the virtual buffers of the relay node provided that the buffers remain stable (i.e., the arrival and departure rates are equal). Due to collision as well as the HD transmission mode at the relay node, the arrival rates at virtual buffers are smaller than $g_1$ and $g_2$, respectively. If the rate of transmission at the relay node becomes lower, the relay node has more opportunities for packet reception which, in turn, increases the throughput. Thus, transmitting coded packets instead of native packets leads to less (one half in symmetric situation) transmission packets, providing more opportunities for packet reception. However, sending a coded packet requires that two constituent packets exist in the virtual buffers which may impose a larger delay in the network. Hence, we are interested in modeling the structure of the relay node by utilizing different transmission probabilities $q_i$ in order to show different tradeoffs among throughput, delay, and transmission power at the relay node. Let us describe the motivation behind the above mechanism. In the case of saturated end nodes, consider a situation that the transmission probability $g_1$ is twice the probability $g_2$; meaning that end node 1 generates more traffic than end node 2. This situation usually happens when there are different traffic categories at the end nodes (similar to different access categories in 802.11e standard) or different traffic densities of the mobile device under the coverage of AP1 and AP2 (see Fig. 1). In these situations, it is possible that sometimes only virtual buffer 1 has packets and virtual buffer 2 is empty. If we now choose different transmission probabilities such as $q_1$ and $q_2$, then, by decreasing $q_1$, node R waits more time to obtain a coding opportunity in order to increase the throughput by sending the coded packets. On the other hand, by decreasing $q_1$, queueing delay at node R increases. Therefore, there seems to be a tradeoff between throughput and queueing delay at node R.

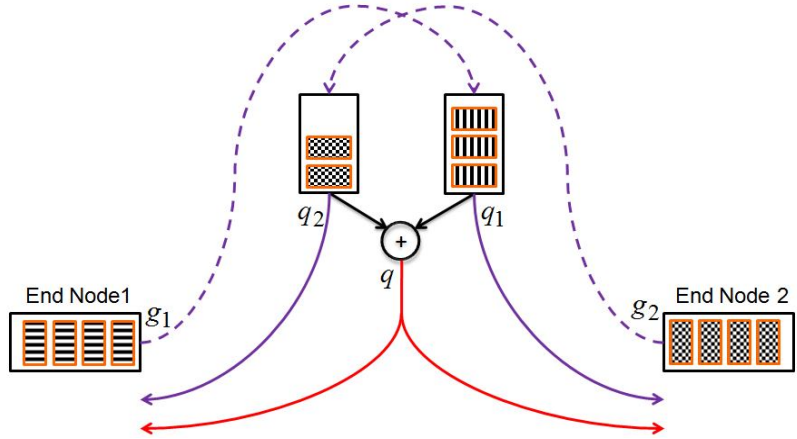

Fig. 2. The Two way relay network structure

A two-way relay network with unsaturated end nodes is shown in Fig. 3, where the relay node R has the same structure as in Fig. 2. Similar to the saturated case, we have assumed that a collision occurs if two end nodes transmit their packets in the same slot. Therefore, emptiness of one end node affects on the departure rate of other end node; that is, both nodes are interacting [16]. Due to different arrival rates $\lambda_1$ and $\lambda_2$ to the end nodes, the buffer of one end node (e.g. end node 2) will probably be less crowded some times. Consequently, the corresponding virtual buffer at the relay node (i.e., virtual buffer 2) may be empty and the chance of having a coding opportunity decreases accordingly. We will show that to deal with this problem, we can decrease the corresponding transmission probability in the relay node (i.e., the transmission probability $q_2$). In this case, we can achieve higher throughput and save some transmission power while the end-to-end delay of each packet increases. In addition, we will be able to find the stability region of the network in several situations.

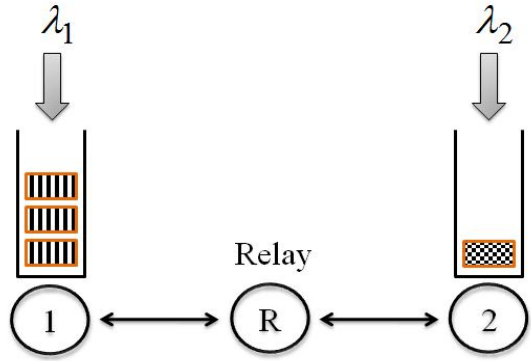

Fig. 3. Two-way relay network with unsaturated end nodes

## III. ANALYSIS OF THE NETWORK WITH SATURATED END NODES

In this section, we introduce our analytical approach to compute throughput, queueing delay, and transmission power in the relay node when both end nodes are saturated. We can model buffer of node R as a Markov chain in which the state $(m, n)$ describes the number of packets in virtual buffers 1 and 2, respectively. The proposed Markov chain is illustrated in Fig. 4, where self-transitions are not shown for convenience. The parameters describing the Markov chain are defined in Table. 1 in which $i, j \in \{1,2\}$.

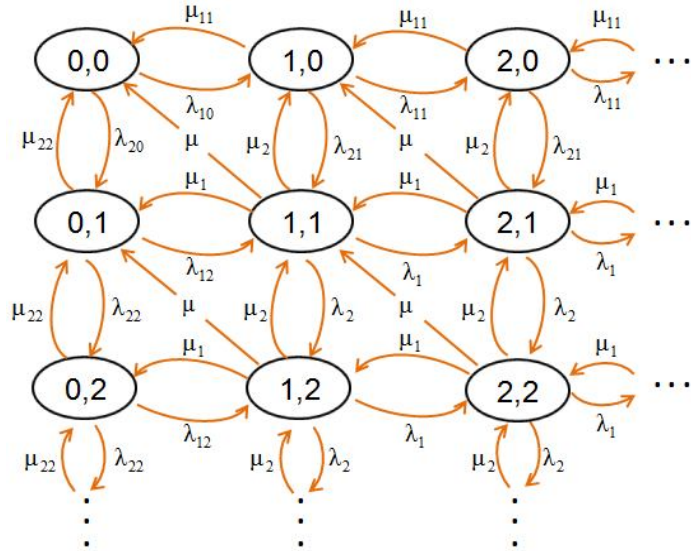

Fig. 4. The Markov chain of the relay node R

Tabel. 1. The parameters describing the Markov chain of node R

| Parameter | Definition |
|---|---|
| $\mu = q(1-g_{\bar{i}})(1-g_j)$ | The probability of having a successful transmission from both virtual buffers when they have packets |
| $\mu_i = q(1-g_{\bar{i}})g_i$ | The probability of having a successful transmission from virtual buffer $i$ when both of them have packets |
| $\mu_{ii} = q_i(1-g_{\bar{i}})$ | The probability of having a successful transmission from virtual buffer $i$ when the other one is empty |
| $\lambda_i = g_i(1-q)(1-g_{\bar{i}})$ | The probability of having a successful reception into virtual buffer $i$ when both of them have packets |
| $\lambda_{ij} = g_i(1-q_j)(1-g_{\bar{i}})$ | The probability of having a successful reception into virtual buffer $i$ when only virtual buffer $j$ has packets |
| $\lambda_{i0} = g_i(1-g_{\bar{i}})$ | The probability of having a successful reception into virtual buffer $i$ when both of them are empty |

Since the end nodes are saturated, by knowing the steady-state probabilities of the above Markov chain, we are able to compute the throughput, i.e., the successful rate of packet transmission. Thus, we need to solve the Global Balance Equation (GBE) for the Markov chain. To this end, we need to first compute the state transition probabilities in the Markov chain representing the relay node R. For example, the state transition probability corresponding to the successful transmission of a native packet is the probability that the native packet is received successfully at its destination. This probability is dependent on the current status of virtual buffers 1 and 2. When both buffers have some packets to transmit, the probability is defined as $\mu_i$ and the corresponding event occurs if node R transmits the coded packet, end node $i$ transmits, and the other end node does not transmit, that is

$$\mu_i = q(1-g_{\bar{i}})g_i. \tag{1}$$

The transition probabilities corresponding to other kind of successful transmissions or receptions can be computed in a similar manner as shown in Table. 1.

Let us define the steady-state probability for the number of packets in the buffers at node R as $\pi(m,n)$, where $m$ and $n$ are the number of packets in virtual buffers 1 and 2, respectively. By solving the mentioned

Markov chain, we are able to derive the steady-state probability $\pi(m,n)$, which is used in the computation of the transmission power, throughput, and queueing delay. The transmitted power in the relay node is proportional to the average number of attempts in a slot made by node R and is expressed as:

$$P = \sum_{m=1}^{\infty} \pi(m,0) q_1 + \sum_{n=1}^{\infty} \pi(0,n) q_2 + \sum_{m=1}^{\infty} \sum_{n=1}^{\infty} \pi(m,n) q \quad . \tag{2}$$

The throughput is the average number of successful transmissions in a slot made by node R and is expressed as:

$$S = \sum_{m=1}^{\infty} \pi(m,0) \mu_1 + \sum_{n=1}^{\infty} \pi(0,n) \mu_2 + \sum_{m=1}^{\infty} \sum_{n=1}^{\infty} \pi(m,n)(\mu_{01} + \mu_{02} + 2\mu) \quad . \tag{3}$$

It worth mentioning that reducing the transmission probabilities $q_i$ have a significant effect on both the transmission power and throughput expressed above, as illustrated in Section IV. This is due to the fact that the number of collisions occurred in the network is effectively influenced by the variations in these transmission probabilities which, in turn, will lead to changes in the transmission power and throughput of the relay node.

By applying Little's Theorem [20], the queueing delay in the buffer at node R is derived as $D = N_R/\lambda_R$, where $N_R$ is the average number of packets in the buffers at node R and $\lambda_R$ is the arrival rate to node R when the system has reached a steady-state. Since node R is stable, the arrival rate $\lambda_R$ is equal to the throughput in (3). Consequently, we have $D = N_R/S$ and $N_R$ is expressed as:

$$N_R = \sum_{m=1}^{\infty} \sum_{n=1}^{\infty} \pi(m,n)(m+n) \quad . \tag{4}$$

In order to find the steady-state probabilities of the Markov chain representing node R, we employ the modeling technique proposed in [19] and utilize distributed Markov chains to represent the status of each virtual buffer. The purpose of this technique, which will be reviewed briefly, is to make two interrelated quasi-birth-death (QBD) processes with a finite number of phases that can be solved analytically. Therefore, we use two Markov chains $MC_1$ and $MC_2$, where $MC_i$ represents the status of virtual buffer $i$ and an estimation of the status of virtual buffer $\bar{i}$. Recursively solving these two Markov chains yields the steady-state probability of node R. For example, as shown in Fig. 5 (self-transition have not been shown), $MC_1$ is a QBD process with $m$ phases and has a set of states $(k_1, \hat{k}_2)$ where $0 \leq k_1$ and $0 \leq \hat{k}_2 \leq m-1$ such that:

$$\hat{k}_2 = \begin{cases} k_2 & ; \quad 0 \leq k_2 \leq m-1 \\ m-1 & ; \quad k_2 > m-1 \end{cases}. \tag{5}$$

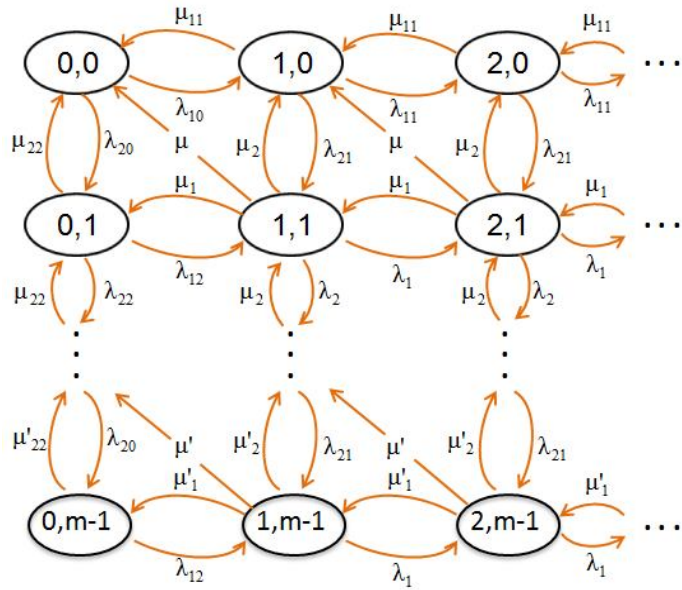

Fig. 5. The distributed Markov chain of virtual buffer 1 ($MC_1$)

Let us define a conditional probability $r_2$ that plays an important role in our analysis. We define this parameter such that:

$$r_2 = \Pr\{k_2 = m-1 | \hat{k}_2 = m-1\}. \tag{6}$$

This probability appears in the state transition probabilities when we are in the last phase of each level of $MC_1$, that is $\hat{k}_2 = m-1$, and we have a successful transmission from virtual buffer 2, as in the following:

$$\mu' = \mu r_2 \quad , \quad \mu'_2 = \mu_2 r_2 \quad , \quad \mu'_{22} = \mu_{22} r_2. \tag{7}$$

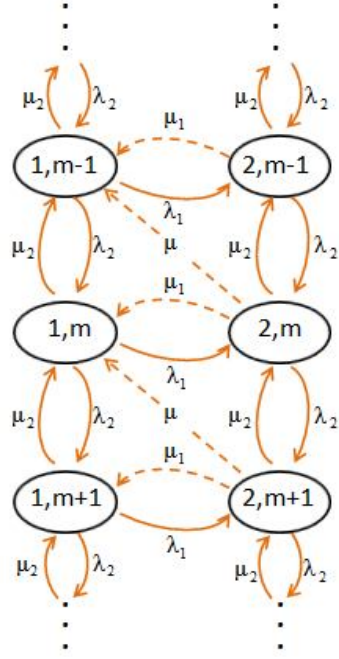

Fig. 6. A part of the original Markov chain of node R

The important change in the transition probabilities belongs to the transition to the previous level at phase $m-1$, named $\mu'_1$. As illustrated in Fig. 6 (the case of $k_1 = 2$), the transition to $(k_1 - 1, m - 1)$ happens in two different corresponding situations in the original Markov chain:

1. When there is a transition from the set of states $\{(k_1, k_2)|\ k_2 > m - 1\}$ to the previous level

2. When there is a transition from the state $(k_1, m - 1)$ to $(k_1 - 1, m - 1)$

Therefore, similar to (7), the corresponding transition probability can be written as:

$$\mu'_1 = (1 - r_2)(\mu + \mu_1) + r_2 \mu_1 = \mu_1 + (1 - r_2)\mu \quad . \tag{8}$$

In a similar manner, the distributed Markov chain of virtual buffer 2 ($MC_2$) is shown in Fig. 7 and the corresponding probabilities are as follow:

$$r_1 = \Pr\{k_1 = m - 1 | \hat{k}_1 = m - 1\} \ ,$$

$$\mu' = \mu r_1 \quad , \qquad \mu'_1 = \mu_1 r_1 \quad , \qquad \mu'_{11} = \mu_{11} r_1 \ ,$$

$$\mu'_2 = (1 - r_1)(\mu + \mu_2) + r_1\mu_2 = \mu_2 + (1 - r_1)\mu \qquad . \tag{9}$$

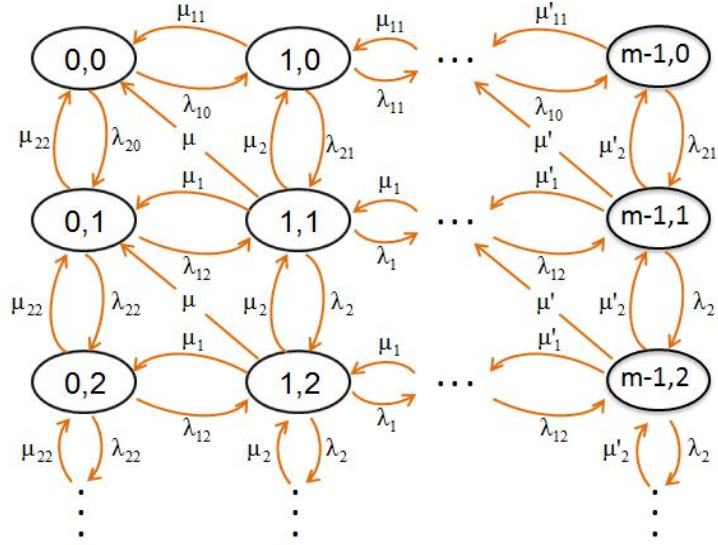

Fig. 7. The distributed Markov chain of virtual buffer 2 ($MC_2$)

Each of these QBD's can be solved by using Matrix Analytic Method introduced in [20]. More detailed illustration of Matrix Analytic Method is included in Appendix A. By applying this method, we propose a simple recursive algorithm that alternates between $MC_1$ and $MC_2$ by updating the conditional probabilities $r_1$ and $r_2$. The algorithm is as follows:

---

**Algorithm 1 – Computing the Conditional Probability**

**INPUT:** Transmission probabilities of the network ($g_1, g_2, q_1, q_2, q$), the number of phases ($m$), the conditional probability $r_i$ in the formation of $MC_j (j \neq i)$, and the arrival rate of the end nodes if we analyze unsaturated case.

**OUTPUT:** conditional probability $r_j$ of $MC_j$

1. By using $r_i$'s ($j \neq i$), form the transition probability matrix of $MC_j$ ($A_{0j}, A_{1j}, A_{2j}, B_{0j}$). (see Appendix A)
2. Compute $R^{(j)}$ from Linear Progression Algorithm
3. Compute $\pi^{(j)}$
4. Set $r_j = \dfrac{\pi^{(j)}_{m-1}}{\sum_{\alpha=m-1}^{\infty} \pi^{(j)}_\alpha}$

---

**Algorithm 2 – Computing the Steady-state Probability**

**INPUT:** Transmission probabilities of the network $(g_1, g_2, q_1, q_2, q)$, the number of phases $(m)$.

**OUTPUT:** the steady-state probability of each MC.

1. Set $k = 0$, set an initial value for the conditional probability $(r_i^0)$.
2. **Repeat**
3. $r_1^{k+1} \leftarrow$ run **Algorithm 1** using $r_2^k$
4. $r_2^{k+1} \leftarrow$ run **Algorithm 1** using $r_1^{k+1}$
5. $k \leftarrow k + 1$
6. **Until** $r_1$ and $r_2$ converge to their steady-state

After setting an initial value for conditional probability $r_2$, we find the steady-state probability vector of virtual buffer 1 in the first iteration, named $\pi^{(1)}$, by using the Matrix Analytic method. On the other hand, we can rewrite (9) as:

$$r_1 = \Pr\{k_1 = m - 1 | k_1 \geq m - 1\} = \frac{\pi_{m-1}^{(1)}}{\sum_{j=m-1}^{\infty} \pi_j^{(1)}} \quad . \tag{10}$$

We then use the new value of $r_1$ for solving the parameters of each $MC$. Then, we can recursively compute the new value of $r_2$ and so on. This process continues until the parameters $r_1$ and $r_2$ converge. As a matter of fact, this is an approximation because each $MC$ can be considered as a truncation of the original Markov chain, but we need to know the matrix $R$ of the whole original $MC$ which due to infinite number of states, is impossible. Our simulations show that this approximation has sufficient accuracy and when these $MC$s are stable, the iteration in **Algorithm 2** always converges. After the convergence, the steady-state probability distribution of the primary Markov chain can be computed. Thereafter, by substituting $\pi(m,n)$ in (2) – (4), we can compute throughput, transmission power, and queueing delay for the relay node.

## IV. NUMERICAL RESULTS FOR SATURATED END NODES

In order to analyze the behavior of a two way relay network, we present some analytical results in different cases. We show that the transmission probabilities of the relay node, $q_1$ and $q_2$, are the critical design parameters of the system to achieve the optimum performance. In our proposed analytical model, we have set $m = 4$ as the number of phases of each distributed Markov chain in the following results. Also, we assume there is an unbalanced generated traffic at the end nodes as $g_1 = kg_2$, where $k$ is called *imbalance factor*. The effect of changing the imbalance factor is illustrated in Fig. 8 in which we have changed the value of $k$ to compare the throughput of the network, while we have already set $q = 0.75$ and $q_1 = q_2 = 0.4$. It can be observed that the more one increases the value of $k$, the more unbalanced traffic are injected into the virtual buffers, and the more one can gain in throughput by reducing the transmission probabilities $q_i$. The last one is due to the fact that reducing the values of $q_i$ relative to $q$ makes the relay's buffers more congested which enables us to have more opportunities to use network coding and therefore, we can gain more throughput. Then, we set $k = 2$ in the following results for convenience; meaning that the offered traffic of end node 1 is twice that of end node 2 and the acceptable range of $g_2$ would be $0 < g_2 < 0.5$.

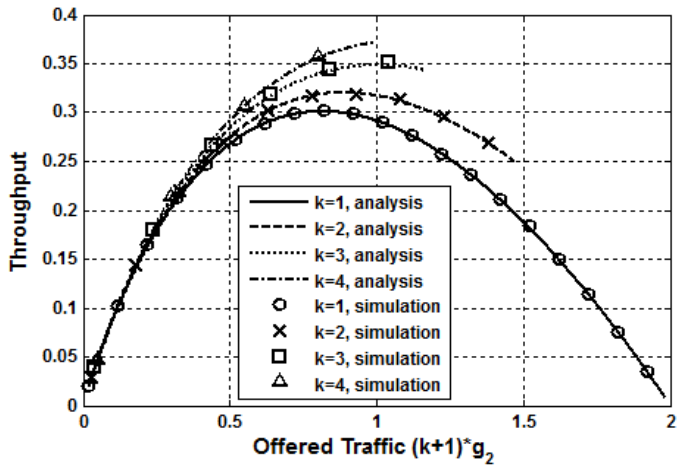

Fig. 8. Throughput versus offered traffic for $q = 0.75$ and $q_1 = q_2 = 0.4$

As a comparison with [15], we refer to Fig. 9, where we have shown the throughput versus the total traffic of the end nodes for different values of $q_1$ and $q_2$. Having set $q = 0.75$, we can observe that by reducing $q_1$ and $q_2$, the throughput of the relay node increases about 8.3% when the offered traffic is high enough. The reason of this behavior will be explained in detail.

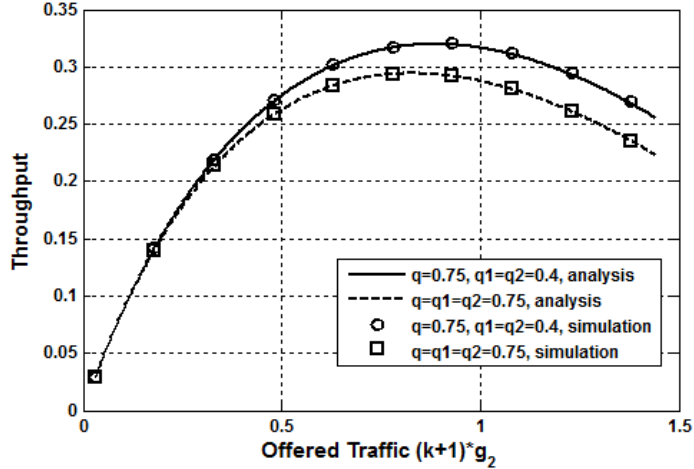

Fig. 9. Offered traffic versus throughput for $k = 2$

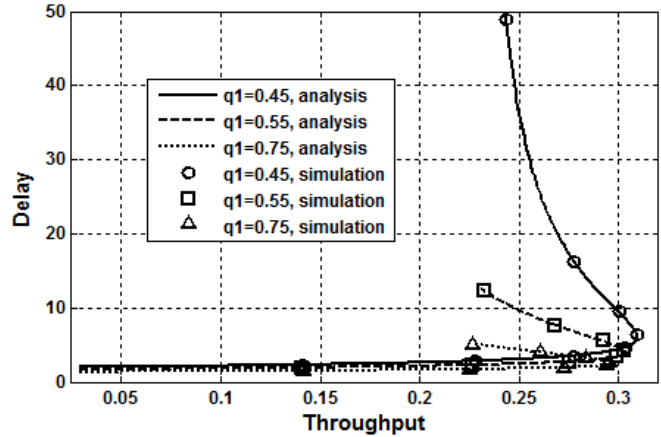

Fig. 10. Delay versus throughput for $q = q_2 = 0.75$, $q_1 = 0.45, 0.55, 0.75$, $0.01 < g_2 < 0.49$

Let us start with inspecting the effect of $q_1$. Changing $q_1$ means changing the transmission probability of the relay node when only virtual buffer 1 has some packets to transmit. For a given set of $(q, q_1, q_2)$, we expect increasing $g_2$ causes an increase in the queueing delay of the relay node. As we observe in Fig. 10, this increment in the queueing delay depends on the value of $q_1$. For a given $q = q_2 = 0.75$, when we decrease $q_1$ to 0.45 in order to achieve more throughput, the queueing delay of the relay drastically increases. As a matter of fact, by decreasing $q_1$, we make the packets in virtual buffer 1 wait longer while there is no packet in virtual buffer 2, in order to obtain a coding opportunity. In this way, we have an increase in throughput because of the lower number of transmissions which requires more time slots to receive the packets (due to HD mode) and therefore, impose more delay in the network. Therefore, we cannot decrease $q_1$ too much because the relay node eventually goes to instability, so there seems to be a tradeoff between throughput and delay of the network. The tradeoff will be more obvious in Fig. 11, where we have shown the variation in delay, throughput, and transmission power of the system for different values of $q_1$ and a given traffic $g_2 = 0.25$ (which corresponds approximately to the peak value in Fig. 9). We observe that $q_1 = 0.3$ is about the optimum value, because any more decrement in $q_1$ extremely increase the queueing delay while almost has no effect on throughput. On the other hand, there is another tradeoff between transmission power and delay, where the transmission power of the relay node continuously decreases along with $q_1$.

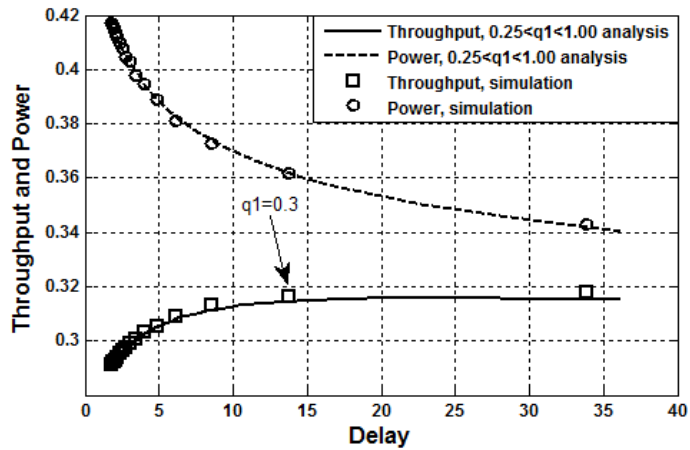

Fig. 11, Throughput and power versus delay for $q = q_2 = 0.75$ and $g_2 = 0.25$

Just like the case of changing $q_1$, we can inspect the effect of $q_2$ in Fig. 12 and Fig. 13. We observe that decreasing $q_2$ does not have the negative effect in queueing delay similar to $q_1$. In fact, when we deal with a traffic with an imbalance factor of $k = 2$, the arrival rate to the virtual buffer 2 is much less than that of virtual buffer 1. Thus, the packets in virtual buffer 2 do not have to wait a long time to get a coding opportunity, therefore, $q_2$ plays a more significant role than $q_1$. Furthermore, the tradeoff between transmission power and queueing delay is again valid and makes us more willing to decrease $q_2$ as much as possible. Therefore, unlike the case of $q_1$, we can decrease $q_2$ in order to achieve a better throughput while the queueing delay of the system remains acceptable.

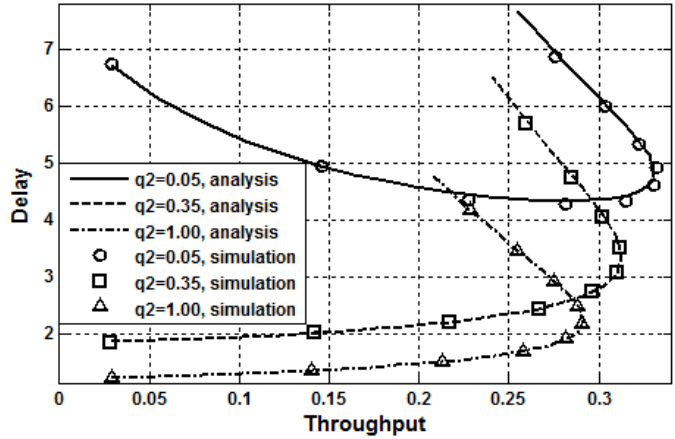

Fig. 12, Delay versus throughput for $q = q_1 = 0.75$ and $0.01 < g_2 < 0.49$

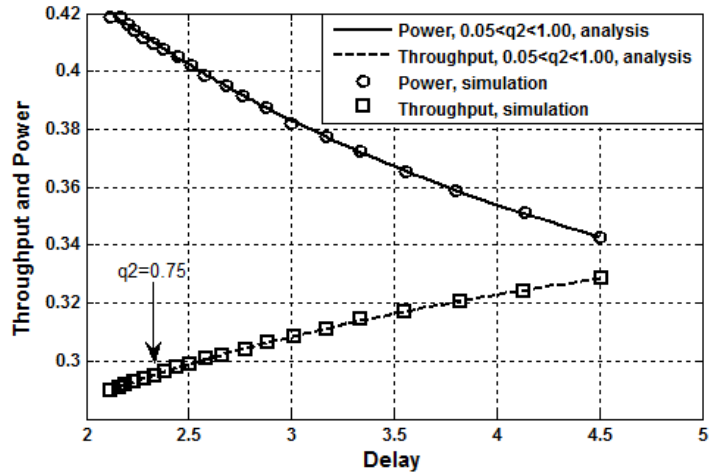

Fig. 13, Throughput and power versus delay for $q = q_1 = 0.75$ and $g_2 = 0.25$

Another interesting observation of Fig. 12 is the behavior of the queueing delay when we set $q_2 = 0.05$. Unlike the case of $q_2 = 0.35$ and $q_2 = 1.00$, where the queueing delay always increases with $g_2$, by setting $q_2 = 0.05$, the queueing delay first experiences a decrement before it starts to go up. This is due to the fact that when the offered traffic of the end nodes ($g_1$ and $g_2$) are both small, the chance of having a coded packet transmission is very low. Increasing $g_2$ provides more opportunities for the relay node to use network coding and transmit probability of $q$ instead of $q_2$ ($q \gg q_2 = 0.05$), which leads to lower waiting time in the relay node's buffer. Therefore, as we offer more traffic to the network, it is expected that the queueing delay decreases.

Finally, we can inspect the effects of changing $q$ in the system performance. As we observe in Fig. 14, decreasing $q$ imposes only more delay on the network while does not change the throughput and transmission power effectively. This behavior is due to the fact that there is no reason to postpone the coding opportunity when both virtual buffers have some packets to transmit. Therefore, decreasing $q$ does not lead to an improvement in throughput, but rather imposes more queueing delay on the network. Hence, we can conclude that in design of a two-way relay network, the most critical parameters are $q_1$ and $q_2$ which can control throughput, delay and transmission power of the relay node.

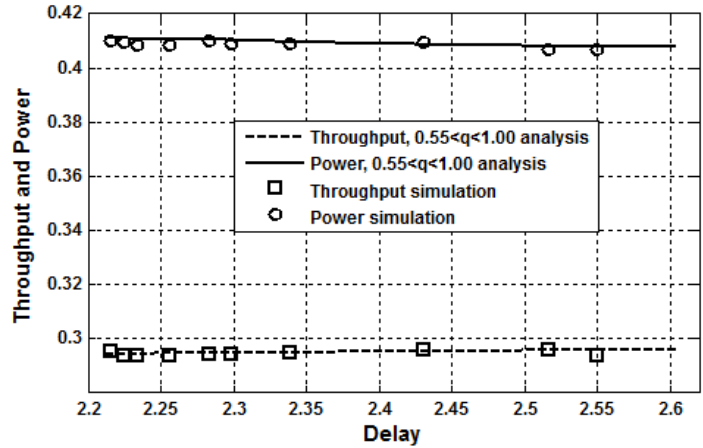

Fig. 14, Throughput and power versus delay for $q_1 = q_2 = 0.75$ and $g_2 = 0.25$

### V. ANALYSIS OF THE NETWORK WITH UNSATURATED END NODES

The case of saturated end nodes is quite similar to the unsaturated one except that the buffers at the end nodes will be empty sometimes. Thus, when we analyze the network with unsaturated end nodes, the arrival rates $\lambda_i$ need to be included in the analysis. To this end, we use the same approach as employed in Section III except that now we have four buffers to deal with. It is worth mentioning that in the case of unsaturated end nodes, the stability region of the network—the set of arrival rates $\lambda_i$'s such that all queues remain

stable—can also be evaluated. By using the proposed analytical model, we are also able to investigate the stability region of the network.

The technique that used in the saturated case can be extended for solving four distributed Markov chains (two Markov chains corresponding to the two virtual buffers and the other two corresponding to end node buffers). Let us define the set of states $(l_1, k_1, k_2, l_2)$ for the Markov chain of the network, where $l_1$ and $l_2$ are the number of packets in the buffers of end nodes 1 and 2, respectively, and $k_1$ and $k_2$ are considered for the two virtual buffers as before. The following formulations is based on the *early departure - late arrival* convention [?]; that is, end node tries to send their existing packets in the beginning of each slot and then receive new packets for transmission. For example, one can see that a transition from state $(l_1, 0, k_2, l_2)$ to state $(l_1, 0, k_2, l_2 + 1)$, $l_1, k_2, l_2 \geq 1$ can occur in three different situations before considering the new packet arrival to the end node's buffer:

1. Both end nodes and node $R$ do not transmit.

2. Both end nodes transmit, thus a collision occurs whether node $R$ transmits or not

3. Node R and end node 1 transmit and end node 2 does not transmit.

After considering new arrivals for the end nodes, the probability of this transition is expressed as:

$$P_{(l_1,0,k_2,l_2) \to (l_1,0,k_2,l_2+1)} = (1-\lambda_1)(1-g_1)(1-q_2)(1-g_2)\lambda_2 + (1-\lambda_1)g_1g_2\lambda_2 + (1-\lambda_1)g_1q_2(1-g_2)\lambda_2. \quad (11)$$

In (11), the indicated transition probability in the Markov chain is not as straightforward as before. On the other hand, the number of different transitions in this Markov chain is much more than we can express here one by one. We describe the details of the transitions in Appendix B.

As an extension of the method employed in Section III, we use four Markov chains $MC_1, MC_2, MC_3$, and $MC_4$ that represent the dynamic of buffer of end node 1, virtual buffer 1, virtual buffer 2, and buffer of end node 2, respectively. For example, $MC_1$ represents the status of the buffer of end node 1 and an estimation of the number of packets in the other buffers. That is to say, $MC_1$ is a QBD process with $m$ phases that has the set of states $(l_1, \hat{k}_1, \hat{k}_2, \hat{l}_2)$, $l_1 \geq 0$ such that:

$$\hat{k}_1 = \begin{cases} k_1 & ; \quad 0 \leq k_1 \leq m-1 \\ m-1 & ; \quad k_1 > m-1 \end{cases}, \qquad \hat{k}_2 = \begin{cases} k_2 & ; \quad 0 \leq k_2 \leq m-1 \\ m-1 & ; \quad k_2 > m-1 \end{cases},$$

$$\hat{l}_2 = \begin{cases} l_2 & ; \quad 0 \leq l_2 \leq m-1 \\ m-1 & ; \quad l_2 > m-1 \end{cases}. \tag{12}$$

The discussed conditional probabilities for each of these QBD's are derived as before. For example, for solving $MC_1$, the following conditional probabilities are necessary:

$$r_2 = \Pr\{k_1 = m-1 | \hat{k}_1 = m-1\}, \qquad r_3 = \Pr\{k_2 = m-1 | \hat{k}_2 = m-1\},$$

$$r_4 = \Pr\{l_2 = m-1 | \hat{l}_2 = m-1\}. \tag{13}$$

These parameters appear in the formulation of the transition probabilities. For example, $r_2$ will be multiplied by the transition probabilities when $\hat{k}_1 = m-1$ and there is a successful transmission from the virtual buffer 1. The formulations of other $MC$'s are straightforward and so we ignore the full details here.

Each of these QBD's can be solved by Matrix Analytic method as done previously. Forming the transition probability matrix $P$ and writing the *Global Balance Equations* are quite similar to the saturated case. The proposed iterative algorithm can be extended as **Algorithm 3**.

**Algorithm 3 – Deriving the Boundary of the Stability Region**

**INPUT:** Transmission probabilities of the network $(g_1, g_2, q_1, q_2, q)$, the number of phases $(m)$.

**OUTPUT:** stability region boundary.

1. Set $l = 0$, set a small value for arrival rate of end node 1 ($\lambda_1^l = 0.01$)
2. **Repeat**
3.     Set $k = 0$, set an initial value for $r_1^0, r_2^0,$ and $r_3^0$.
4.     Set conditional probability of $MC_4$ zero ($r_4 = 0$)
5.     **Repeat**
6.         $r_1^{k+1} \leftarrow$ run **Algorithm 1** using $r_2^k, r_3^k, r_4,$ and $\lambda_1^l$.
7.         $r_2^{k+1} \leftarrow$ run **Algorithm 1** using $r_1^{k+1}, r_3^k, r_4,$ and $\lambda_1^l$.
8.         $r_3^{k+1} \leftarrow$ run **Algorithm 1** using $r_1^{k+1}, r_2^{k+1}, r_4,$ and $\lambda_1^l$.
9.         $k \leftarrow k + 1$
10.     **Until** $r_1, r_2,$ and $r_3$ converge to their steady-state probabilities.
11.     Set $t = 0$, choose a reasonable value for $\lambda_2^t$
12.     **Repeat**
13.         $r_4^{k+1} \leftarrow$ run **Algorithm 1** using $r_1, r_2, r_3, \lambda_1^l,$ and $\lambda_2^t$.
14.         $t \leftarrow t + 1$
15.         $\lambda_2^{t+1} \leftarrow \lambda_2^t + 0.001$
16.     **Until** end node 2 is saturated (means that $r_2$ is very close to zero.)
17.     The pair of $(\lambda_1^l, \lambda_2^t)$ is a point on the boundary of the stability region
18.     $l \leftarrow l + 1$
19.     $\lambda_1^{l+1} \leftarrow \lambda_1^l + 0.001$
20. **Until** end node 1 is saturated (means that $r_1$ is very close to zero.)

In our numerical analysis, we select transmission probabilities in the network such the relay node does not saturate. Therefore, we intuitively set values for the transmission probabilities $q, q_1,$ and $q_2$ which are larger than $g_1$ and $g_2$. In **Algorithm 3**, we set a small value for the arrival rate at end node 1 (e.g. $\lambda_1 = 0.01$) and find the largest arrival rate $\lambda_2$ that saturates end node 2. Then, we increase the value of $\lambda_1$ by a small step (e.g. step $= 0.001$) and so forth. In order to derive the queueing delay, transmission power, and throughput of the network in the unsaturated case, we can do the same as Algorithm 2 except that now our iterative algorithm should solve four MCs to obtain the conditional probabilities in (13).

## VI. NUMERICAL RESULTS FOR UNSATURATED END NODES

In order to verify that the proposed Algorithm is able to derive the stability region of the network, we run our algorithm in deriving the stability region for some parameter sets and compare the results with simulations. We first show the **Algorithm 3** can accurately specify the boundary of the stability region, then, we present a detailed comparison of network performance in different cases—regarding the delay, transmission power, and stability region.

The stability region of the network is shown in Fig. 15, where the imbalanced factor of the network is set to $k = 1$. The non-NC case is the mechanism in which the relay node does not combine the packets of the two virtual buffers and only transmits the native packets to their destinations with a single transmission probability $q$. As illustrated in Fig. 15, utilizing the network coding mechanism—even by using the same transmission probability $q = q_1 = q_2 = 0.7$—improves the stability region; because sending a coded packet helps the network to empty out its buffers faster. Furthermore, simultaneously reducing $q_1$ and $q_2$ leads to an expansion in the stability region; the more balanced the incoming traffic is, the more distinct the improvement in the stability region is. This is because of the fact that the network is more congested in this set of parameters and therefore, utilizing the network coding will be of great help in handling the congestion.

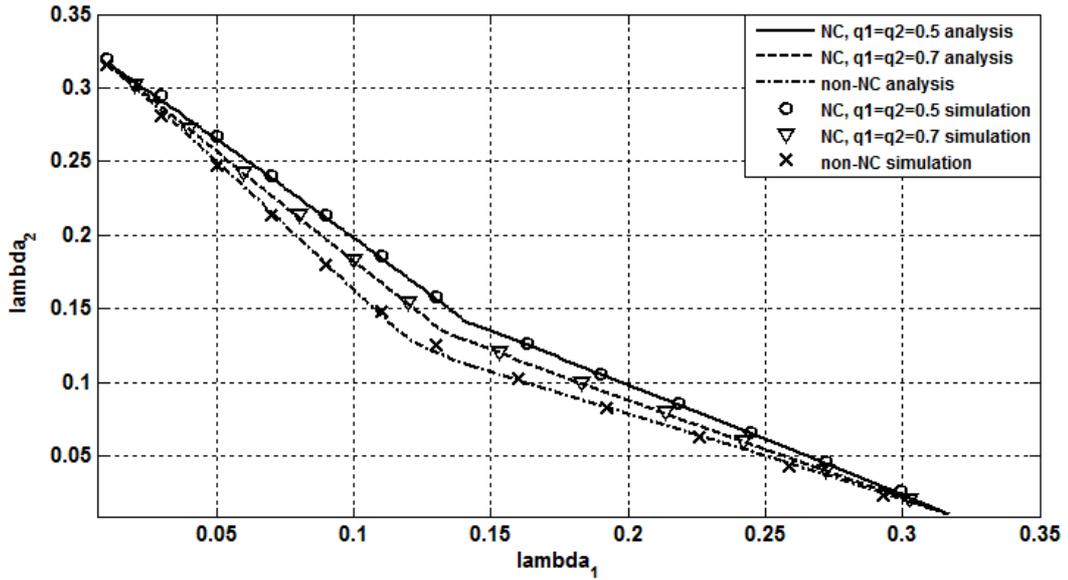

Fig. 15, The stability region boundary of the network, $q = 0.7, g_2 = 0.5, k = 1, m = 6$

This approach enables us to achieve the stability region of the network in several situations. For example, when the imbalanced factor is set to $k = 2$, the stability region is relatively different compared to the previous one, as shown in Fig. 16. Here, the network parameters have been chosen such that the differences between the three scenarios become more obvious. Another example of stability region is shown in Fig. 17, where, unlike Fig. 16, the difference between non-NC scenario and NC scenario with $q = q_1 = q_2$ is negligible. In this case, the reasonable choice to expand the stability region efficiently is to reduce the transmission probabilities $q_i$, because here, the end node buffers are less crowded than that in Fig. 16, and therefore, the relay node can wait more to obtain a coding opportunity, while does not induce a large delay in the network. The point we are trying to make here is, regardless of the network parameters, reducing the transition probabilities $q_i$ is efficient in extending the stability region of the network. However, similar to the saturated case in Section IV, this reduction is not a perfect solution and might have some negative effects in the network's performance such as delay increament which is illustrated in the following.

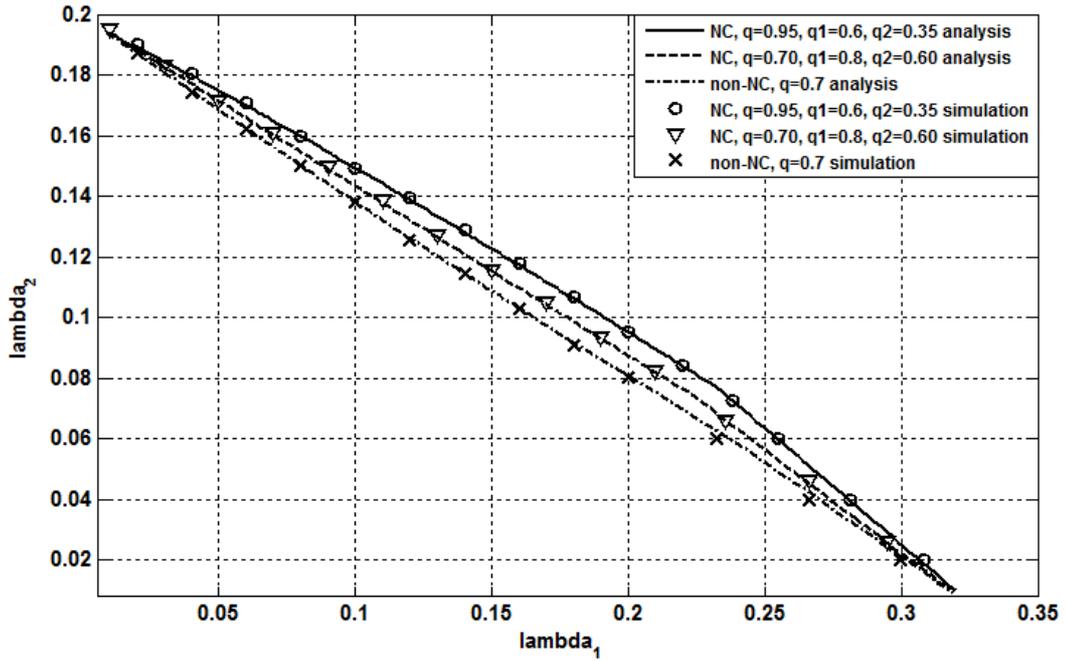

Fig. 16, The stability region boundary of the network, $g_2 = 0.25, k = 2, m = 6$

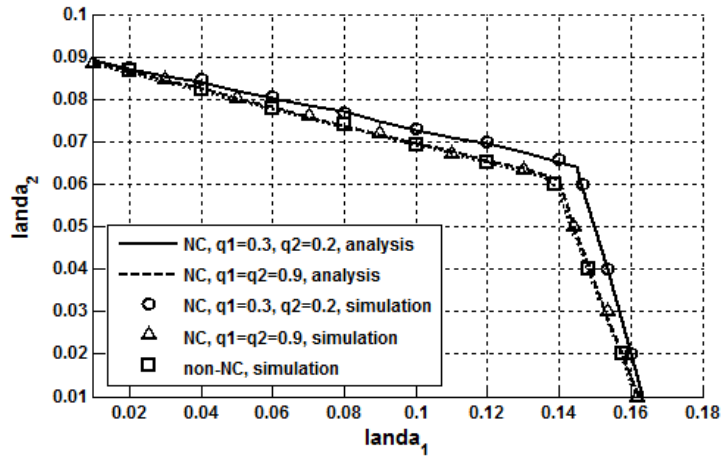

Fig. 17. The stability region boundary of the network, $q = 0.9, g_2 = 0.1, k = 2, m = 6$

Let us start with inspecting the behavior of the queueing delay in the network. The queueing delay of the relay node in three different scenarios is shown in Fig. 18, where the network parameters are those employed in Fig. 15. As expected, it can be observed that by reducing $q_i$, the network deals with an increment in the queueing delay which is the price to pay for extending the stability region. In addition, the fact that this reduction leads to an expansion in the stability region—as it happened in Fig. 15—is usually of more importance than the small imposed queueing delay; which is why we are interested in utilizing this new mechanism rather than the one with $q = q_1 = q_2$. Furthermore, the transmission power of the relay node is improved by these modifications which is illustrated is Fig. 19; where reducing $q_i$ obviously makes the best performance of all three.

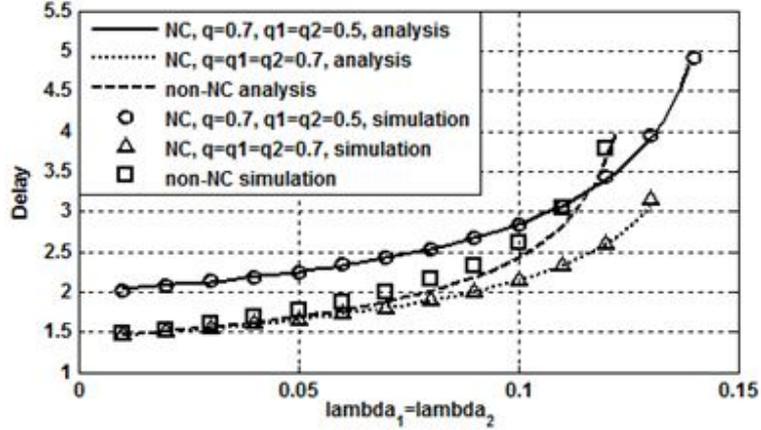

Fig. 18. Delay versus the arrival rate $q = 0.7$, $\lambda_1 = \lambda_2$, $g_1 = g_2 = 0.5$

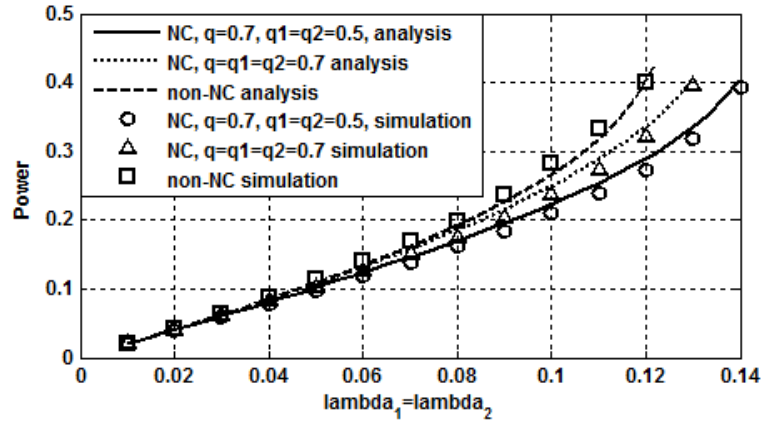

Fig. 19. Transmission power versus the arrival rate $q = 0.7$, $\lambda_1 = \lambda_2$, $g_1 = g_2 = 0.5$

## VII. Conclusion

We showed that, in a two-way relay network with saturated or unsaturated end nodes, we can use network coding to exchange our packets in a more effective way than the one proposed in previous works. This advantage was more evident when there is an unbalanced offered traffic in the end nodes. We analyzed the network using three different transmission probabilities and showed that the critical design parameters in such a network are the transmission probabilities of sending the native packets; that is, when only one virtual buffer has some packets to transmit and the other one is empty. We could design these parameters to achieve more throughputs while remaining within a reasonable range for queueing delay. Additionally, by extending this new approach, we obtained the stability region boundary of TWRN in different situations when the end nodes are unsaturated.

## APPENDIX A

In general, a QBD process is a Markov comprised of states $\{(l,i)|l \geq 0, 1 \leq i \leq m\}$, where the state space can be divided into levels, and each level $l$ has $m$ states (phases). In a QBD process, transitions are allowed only to the neighboring levels or within the same level. When transition probabilities between levels—except for those from and within the first level—are alike, the QBD is said to be homogeneous. Thus, a homogeneous QBD process has a transition probability matrix expressed as

$$P = \begin{pmatrix} B_1 & A_0 & 0 & 0 & \ldots \\ A_2 & A_1 & A_0 & 0 & \ldots \\ 0 & A_2 & A_1 & A_0 & \ldots \\ 0 & 0 & A_2 & A_1 & \ldots \\ \vdots & \vdots & \vdots & \vdots & \ddots \end{pmatrix}, \tag{14}$$

where $A_0, A_1, A_2$, and $B_1$ are square matrices containing the transitions between the corresponding states. Specifically, the inner sub-matrices of $P$ are as follow:

$$B_1 = P_{(0,:) \to (0,:)},$$

$$A_0 = P_{(k,:) \to (k+1,:)}, \quad k \geq 0,$$

$$A_1 = P_{(k,:) \to (k,:)}, \quad A_2 = P_{(k,:) \to (k-1,:)}, \quad k \geq 1. \tag{15}$$

Let us show the steady-state probability of this QBD by $\pi = (\pi_0, \pi_1, \pi_2, \ldots)$, where $\pi_i = (\pi_{i,0}, \pi_{i,1}, \pi_{i,2}, \ldots, \pi_{i,m-1})$, $0 \leq i \leq \infty$ is the steady-state probability vector of level $i$ of the QBD. Then, the *Global Balance Equations* can be written as

$$\pi_0(B_1 - I) + \pi_1 A_2 = 0,$$

$$\pi_0 A_0 + \pi_1(A_1 - I) + \pi_2 A_2 = \pi_0 A_0 + \pi_1(A_1 - I + RA_2) = 0,$$

$$\pi_0 \mathbf{1} + \pi_1(I - R)^{-1}\mathbf{1} = 1 , \qquad (16)$$

where we have used the normalization condition $\pi \mathbf{1} = 1$ and the following matrix geometric property:

$$\pi_i = \pi_{i-1} R = \pi_1 R^{i-1} \quad , \quad i \geq 1 . \qquad (17)$$

$R$ is a square matrix similar to $A_i$ such that $R_{ij}$ indicate the average number of visits of phase $j$ at $(n+1)^{th}$ level between any two consecutive visits of $n^{th}$ level conditioned on the first visit of $n^{th}$ level be at $i^{th}$ phase. Therefore, we first need to compute the rate matrix $R$ used in (17). There are different ways to find the Rate Matrix $R$ based on characterization of the problem. We have employed "Linear Progression Algorithm" introduced in [21]. Now, we can easily solve (16) to obtain the steady-state probability of the QBD.